\documentclass[aps,prl,amsfonts,amssymb,twocolumn,amsmath,floatfix,showpacs]{revtex4-1}
\usepackage{amsbsy}
\usepackage{subfigure}
\usepackage[dvips]{color}
\usepackage[dvips]{graphics}
\usepackage{graphicx}
\usepackage{bm}

\newcommand{\bel}{\begin{align}}
\newcommand{\bem}{\begin{multline}}
\newcommand{\beq}{\begin{equation}}
\newcommand{\bea}{\begin{eqnarray}}
\newcommand{\eeq}{\end{equation}}
\newcommand{\eea}{\end{eqnarray}}
\newcommand{\eel}{\end{align}}
\newcommand{\eem}{\end{multline}}

\begin{document}

\title{Non-Fraunhofer patterns of the anharmonic Josephson
current influenced by a strong interfacial pair breaking}

\author{Yu.\,S.~Barash}

\affiliation{Institute of Solid State Physics, Russian Academy of Sciences,
Chernogolovka, Moscow District, 142432 Russia}


\begin{abstract}
In the junctions with a strong Josephson coupling and a pronounced
interfacial pair breaking, the magnetic interference patterns of the
Josephson current are shown to differ substantially from the standard
Fraunhofer shape. The Fraunhofer pattern occurs, when Josephson
couplings are weak. The narrow peak of the critical current,
centered at the zero magnetic field, and the suppressed hills at
finite field values are the characteristic features of the
non-Fraunhofer magnetic field modulation of the critical current,
obtained in this paper.
\end{abstract}


\maketitle

The Josephson critical current, as a function of the applied
magnetic field, manifests a characteristic modulation, which
originates from the fundamental quantum coherence of the
superconducting leads. Under standard conditions the modulation is
described by the Fraunhofer diffraction pattern, which has
represented a benchmark for the magnetic field modulation of the
critical current since the early days of the discovery of the
Josephson effect. \cite{Rowell1963,Josephson1964,Barone1982,%
Tinkham1996} Distortions of the Fraunhofer shape in tunnel junctions
stem from inhomogeneities or from variations of geometry of the
fabricated interfaces. \cite{Barone1982,Tinkham1996} In contemporary
studies of the magnetic field induced interference patterns, single
junctions are usually fabricated as a sequence of different regions,
e.g., of $0\,$ and $\pi\,$ regions. In such junctions the patterns
can substantially deviate from the standard shape, due to their
phase sensitivity. The corresponding measurements constitute the
basis of the Josephson interferometry method \cite{VanHarlingen1995,%
VanHarlingen1995_2,tsueikirt00,Mannhart2002,Hilgenkamp2008,%
DellaRocca2005,Goldobin2006,Nelson2004,VanHarlingen2006,Halperin2009,%
Ryazanov2006,Goldobin2010,Mints2010,Linder2012}, which has been
successfully used to identify symmetries of, e.g., the $d$-wave
order parameters of cuprate superconductors
\cite{VanHarlingen1995_2,VanHarlingen1995,tsueikirt00}, the complex
$p$-wave superconductivity in $\text{Sr}_2\text{RuO}_4$
\cite{Nelson2004,VanHarlingen2006} and the $E_{2u}$ pairing of the
superconducting $\text{UPt}_3$ \cite{Halperin2009}. Single Josephson
junctions with inhomogeneous magnetic interlayers, fabricated as a
sequence of $0\,$ and $\pi\,$ regions, also exhibit
the interference patterns, which qualitatively differ from the
Fraunhofer ones. \cite{Ryazanov2006,Goldobin2010,Mints2010,%
Linder2012}

This paper theoretically studies single junctions with homogeneous
thin interfaces and focuses on the effects of the anharmonic phase
dependence of the Josephson current on the magnetic field induced
modulation. The modulation of the Josephson current will be described
within the Ginzburg-Landau (GL) theory. The effects will be shown to
become pronounced in the junctions with a strong Josephson coupling,
where the modulation differs heavily from the Fraunhofer diffraction
pattern. The results obtained could be, in particular, implemented in
the junctions involving unconventional superconductors and/or
magnetic interfaces. An intensive interfacial pair breaking is
crucial for the systems in question. It substantially suppresses the
critical current, therefore maintaining the planar junctions with a
pronounced Josephson coupling as the weak links with strongly
anharmonic current-phase relations. \cite{Barash2012}

\begin{figure}[!thb]
\centering
\includegraphics[width=1.\columnwidth,clip=true]{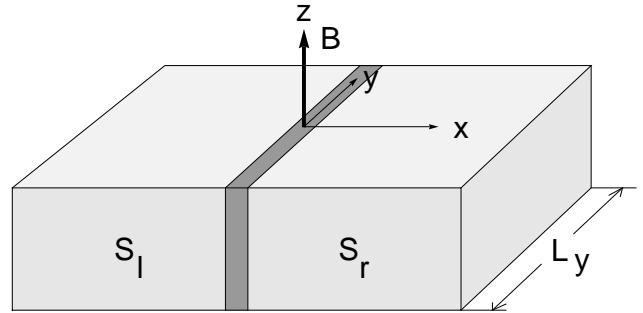}
\caption{Schematic diagram of the junction.}
\label{scheme}
\end{figure}

Assume the usual form of the GL free energy, which applies,
for example, to $s$-wave and $d_{x^2-y^2}$-wave junctions. Let the
symmetric junctions have a spatially constant width much less than
the Josephson penetration length. The superconducting electrodes
${\cal S}_l$ and ${\cal S}_r$ are supposed to be thick compared to
the magnetic penetration depth $\lambda$, while a homogeneous
plane rectangular interlayer at $x=0$ is assumed to be of zero length
within the GL approach (see Fig.\,\ref{scheme}). The interface
contribution to the GL free energy incorporates both the Josephson
coupling of the two superconducting banks $g_J|\Psi_+ -\Psi_-|^2$
and the term $g(\left| \Psi_+\right|^2+\left|\Psi_-\right|^2)$,
which is responsible for the interfacial pair breaking or pair
producing. It will be convenient to use the dimensionless Josephson
coupling constant $g_\ell=g_J\xi(T)/K$, the interface pair
breaking parameter $g_{\delta}=g\xi(T)/K$ and their combination
$g_b(\chi)=(g_{\delta}+2g_\ell\sin^2\frac{\chi}2)$, where $\xi(T)$
is the temperature dependent superconducting coherence length, $K$
comes from the bulk gradient term $K|\nabla\Psi|^2$ and $\chi$
is the phase difference of the order parameters across the interface.

Microscopic estimations show that the range of variations of the
parameters $g_{\ell}$ and $g_{\delta}$ (and, therefore, of $g_b$) is
quite wide and, in particular, includes large values.
\cite{Barash2012,Barash2012_2} The limit $g_{\delta}\to+\infty$
corresponds to the order parameter vanishing at the boundary.
In highly transparent junctions ($(1-{\cal D})\ll 1$) the
parameter $g_{\ell}\propto(1-{\cal D})^{-1}$ can be arbitrarily large.
Here ${\cal D}$ is the transmission coefficient. For small and
moderate transparencies in dirty $s$-wave junctions $g_\ell$ can
vary from vanishingly small values in the tunneling limit to those
well exceeding $100$ near $T_c$.  For the pair breaking surfaces
($g_{b}>0$) arbitrarily large values of
$g_b$ are admissible in the GL theory, since they correspond
to spatial variations of the order parameter on a scale
$\agt\xi(T)$. This is in contrast to the pair producing interfaces
$g_b<0$, where the superconductivity is locally enhanced (see, e.g.,
\cite{Fink1969,Buzdin1987,Geshkenbein1988,Samokhin1994,Indekeu2007,%
Barash2012_2}). Near the boundary a strong enhancement can induce the
scale substantially less than the coherence length of the leads.

The self-consistent description of the Josephson current in planar
junctions, recently developed within the GL theory
in Ref.\,\onlinecite{Barash2012}, takes into account the pair breaking
effects produced by the phase-dependent Josephson coupling as well as
by the current and by the interface itself. In particular, the following
anharmonic current-phase relation has been obtained in the absence of the
applied magnetic field:
\begin{multline}
j\left(g_{\ell},g_{\delta},\chi\right)=
\frac{3\sqrt{3}g_{\ell}\sin\chi}{2(1+2g_{\ell}^{2}
\sin^2\chi)}\biggl[1+g_b^{2}(\chi)+
g_{\ell}^{2}\sin^2\chi-\\
-\sqrt{\bigl(g_b^{2}(\chi)+g_{\ell}^{2}\sin^2\chi\bigr)^2+
2g_b^{2}(\chi)}\,\biggr]j_{\text{dp}}.
\label{bcss1012p}
\end{multline}
Here $j_{\text{dp}}$ is the depairing current in the bulk.

Equation \eqref{bcss1012p} describes the current behavior almost
perfectly, if $j< 0.7j_{\text{dp}}$. This concerns, in particular,
the current at $g_{\ell}<1$ for any $g_{\delta}$, or at $g_{\delta}>
1$ for any $g_{\ell}$. For $j>0.7j_{\text{dp}}$ equation
\eqref{bcss1012p} reasonably approximates the exact numerical
solution, with the deviations not exceeding $10\%$.
\cite{Barash2012} The solution \eqref{bcss1012p} applies only to the
pair breaking interfaces, i.e., to $g_b(\chi)\ge0$. Otherwise, the
parameters in equation \eqref{bcss1012p} can take arbitrary values.
In order to describe the current flowing through the
pair producing interface, one should change the sign before the
square root in equation \eqref{bcss1012p}. Depending on $g_{\ell}$ and
$g_{\delta}$, such current could exceed $j_{dp}$ and, hence, destroy
the superconductivity in the bulk of the leads. From now on only the pair
breaking interfaces with $g_{\ell}, g_{\delta}>0$ will be considered.

For a pronounced interfacial pair breaking $g_{\delta}^2\gg 1$
expression \eqref{bcss1012p} for the supercurrent is simplified and
reduced to
\begin{equation}
j\approx\frac{3\sqrt{3}g_\ell\sin\chi}{4[
g_\delta^2+4(g_\delta+g_\ell)g_\ell
\sin^2\frac{\chi}2]}j_{\text{dp}}.
\label{cpr3}
\end{equation}
The corresponding critical current $j_{\text{c}}=[3\sqrt{3}g_\ell/
4g_\delta(g_\delta+2g_\ell)]j_{\text{dp}}\ll j_{\text{dp}}$ is
always small, at arbitrary $g_{\ell}$. The validity of the condition
$j\ll j_{\text{dp}}$ is the evidence of a weak link. Strongly anharmonic
current-phase relation shows up in \eqref{cpr3} for
$g_\ell^2\gg g_\delta^2\gg 1$, while in the case $g_\ell \ll
g_{\delta}$ the first harmonic dominates the current.

Now let the magnetic field be applied to the junction along the $z$
axis:\,\,$\bm{B}(x)=B(x)\bm{e}_z$ (see Fig.~\ref{scheme}). It is
assumed that the field is not too strong, substantially less than
the critical fields of the leads. The well known gauge invariant
result states that the supercurrent density in the presence of the
field is described by its zero-field expression $j(\chi)$, if the
spatially dependent phase difference across the interface
$\widetilde\chi(y)=\chi+2\pi(y/L_y)(\Phi\big/\Phi_0)$, influenced
by the magnetic flux $\Phi$ through the junction, is substituted for
$\chi$. \cite{Barone1982,Tinkham1996} Here  $\Phi_0=\pi\hbar c/|e|$
is the superconducting flux quantum. Considering the total
supercurrent through the junction is of interest, the spatially
modulated supercurrent density should be averaged over the interface
area:
$
I=\frac{1}{L_y}\!\int_{-\frac{L_y}{2}}^{\frac{L_y}{2}}\!
j\left(\chi+\frac{2\pi y\Phi}{L_y\Phi_0}\right) dy=
\frac{\Phi_0}{2\pi\Phi}\int_{\varphi_-}^{\varphi_+}\!
j(\varphi) d\varphi
$.
Here a rectangular plane interface is supposed to occupy the space
$(-L_y/2, L_y/2)$ along the $y$ axis. The integration limits
$\varphi_{\pm}$ are defined as $\varphi_{\pm}=\chi\pm\pi(\Phi\big/
\Phi_0)$.

\begin{figure*}[!thb]
\centering
\subfigure[\,\,\, tunnel junctions\,\,\, ($g_{\ell}\ll 1$) \label{0} ]{\includegraphics[width=.78\columnwidth,clip=true]{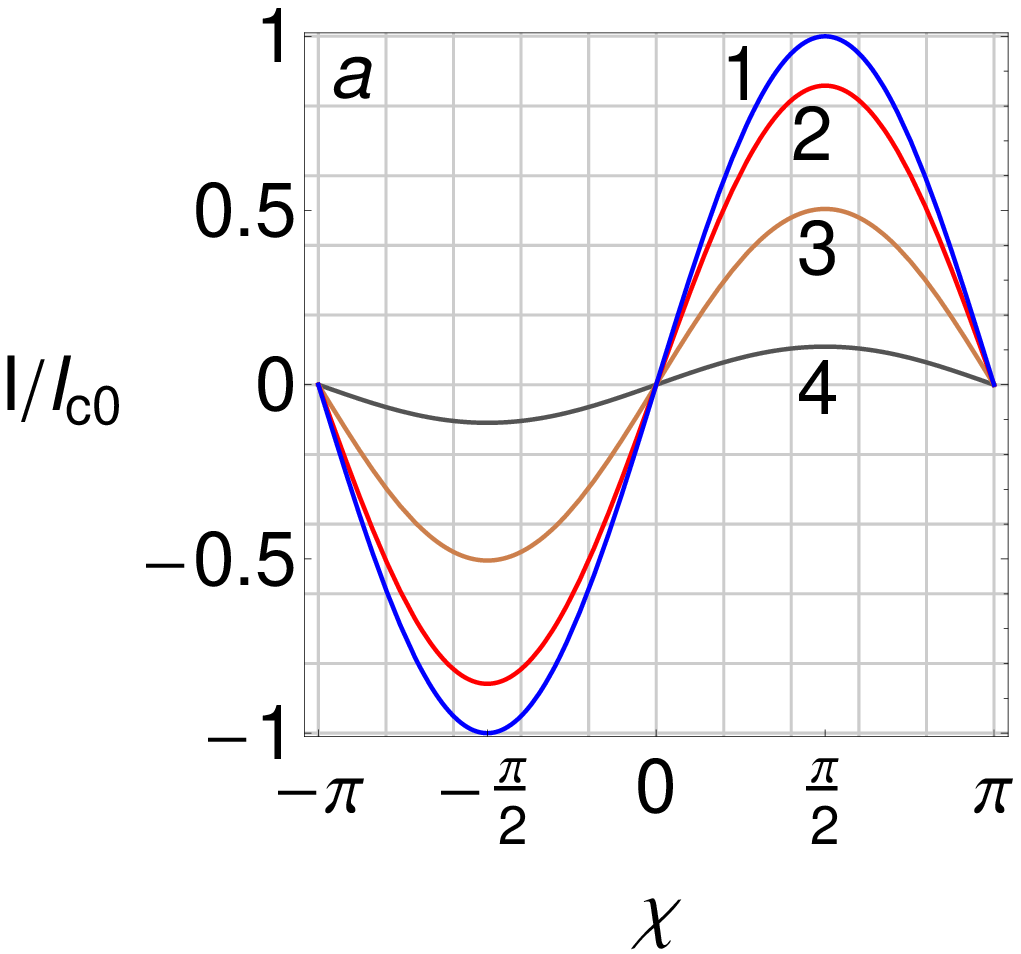}}
\subfigure[\,\,\, $g_{\ell}=10$ \vspace{0.01cm}\label{1}]{\includegraphics[width=.572\columnwidth,clip=true]{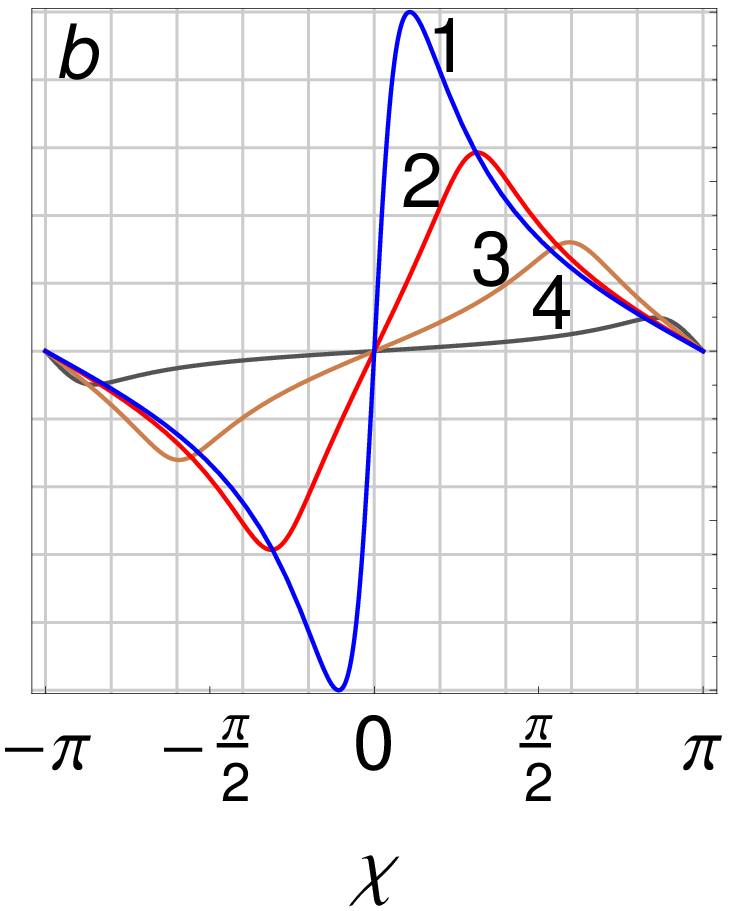}}
\subfigure[\,\,\, $g_{\ell}=100$ \label{2}]{\includegraphics[width=.699\columnwidth,clip=true]{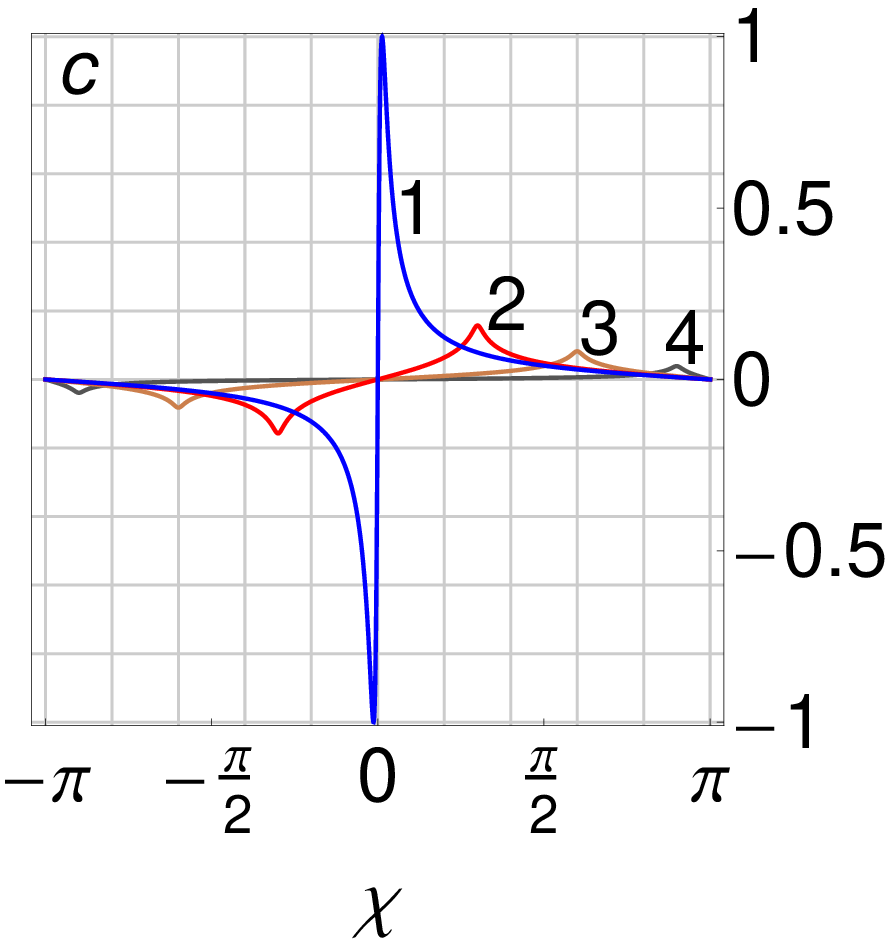}}
\caption{The current-phase relations in junctions with the pair
breaking parameter $g_{\delta}=4$, taken at several values of the
magnetic flux $\Phi$: (1)\, $\Phi=0$\,\, (2)\, $\Phi=0.3\Phi_0$\,\,
(3)\, $\Phi=0.6\Phi_0$\,\, (4)\, $\Phi=0.9\Phi_0$\,. Here
$\Phi_0=\pi\hbar c/|e|$ is the flux quantum. Different panels
correspond to different Josephson coupling constants. For each
coupling constant the current is normalized to its critical value in
the absence of the field.}
\label{cprs}
\end{figure*}

A macroscopic scale of the modulation period $L_y^B=\pi\ell_B^2/2
\lambda=\Phi_0/2B(0)\lambda$ allows one to consider, for a given $y$,
local densities per unit interface area of thermodynamic potential
and of the supercurrent, which satisfy the relation $j(\chi)=
\frac{2|e|}{\hbar}\frac{d}{d\chi}{\mathit\varOmega}_0(\chi)$. Here
$\ell_B=(\hbar c/|eB(0)|)^{1/2}$ is the magnetic length. With this
relation, the Josephson current averaged over the interface is
\cite{Barash2010}
\begin{equation}
I=\frac{|e|\Phi_0}{\pi\Phi\hbar}\left[{\mathit\varOmega}_0\Bigl(\chi+
\frac{\pi\Phi}{\Phi_0}\Bigr)-
{\mathit\varOmega}_0\Bigl(\chi-\frac{\pi\Phi}{\Phi_0}\Bigr)\right].
\label{I_varOmega0}
\end{equation}
Thus, the phase dependent thermodynamic potential of the junction in
the absence of the field also describes the magnetic interference
pattern in the junction.

Equation \eqref{I_varOmega0} will be now used to establish the
modulation of the anharmonic Josephson current by the applied
magnetic field. Interestingly, for a strong interfacial
pair breaking $g_{\delta}^2\gg1$, when the supercurrent in
the absence of the field is described by \eqref{cpr3}, the averaged
anharmonic current allows not only a numerical study but also
an analytical description. Indeed, integrating \eqref{cpr3}
over the phase difference and using \eqref{I_varOmega0}, one obtains
the magnetic field dependent current-phase relation for the averaged
anharmonic supercurrent:
\begin{multline}
I(\chi,\Phi)=j_{\text{dp}}\frac{\Phi_0}{\pi\Phi}\dfrac{3\sqrt{3}}{16
\left(g_{\delta}+g_{\ell}\right)}\times\\ \times\ln\!\left[1+
\dfrac{2\sin\chi\sin\left(\frac{\pi\Phi}{\Phi_0}\right)}{
\frac{g_{\delta}^2}{2\left(g_{\delta}+g_{\ell}\right)g_{\ell}}
+2\sin^2\bigl(\frac{\chi}2-\frac{\pi\Phi}{2\Phi_0}\bigr)}\right].
\label{IvarOmega03}
\end{multline}

The corresponding critical current is
\begin{equation}
I_c\!=\!\frac{3\sqrt{3}j_{\text{dp}}\Phi_0}{16\pi|\Phi|\left(g_{\delta}+g_{\ell}\right)}
\ln\biggl[1+\dfrac{2A}{\sqrt{A^2+
g_{\delta}^2\left(g_{\delta}+2g_{\ell}\right)^2}-A}\Biggr],
\label{IvarOmega20p}
\end{equation}
where $A=2\left(g_{\delta}+g_{\ell}\right)g_{\ell}
\left|\sin\left(\frac{\pi\Phi}{\Phi_0}\right)\right|$.

\begin{figure}[b]
\includegraphics[width=\columnwidth,clip=true]{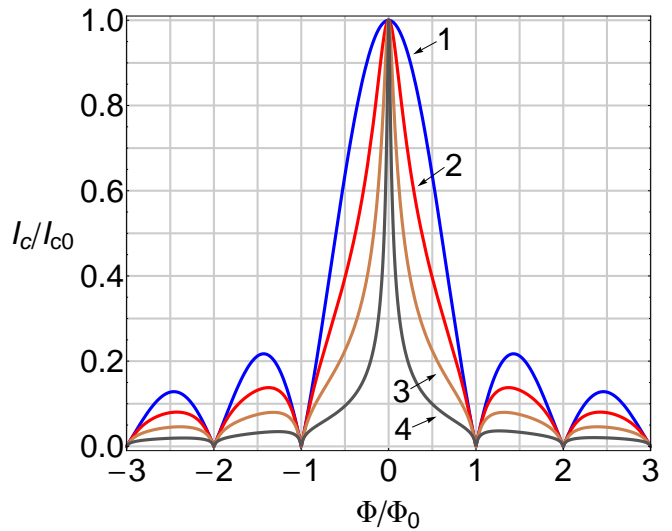}
\caption{The normalized critical current
$I_{\text{c}}\big/I_{\text{c}0}$ as a function of the magnetic flux,
which pierces the junctions with different Josephson coupling
constants:\,\, (1)\, small Josephson couplings $g_{\ell}\ll 1$,
described by the Fraunhofer pattern\,\, (2)\, $g_{\ell}=10$\,\, (3)\,
$g_{\ell}=30$\,\, (4)\, $g_{\ell}=100$. The interfacial pair breaking
parameter is  $g_{\delta}=4$; $I_{\text{c}0}$ is the zero-field
critical current for the corresponding junctions. The curves have
been plotted based on equations \eqref{bcss1012p} and
\eqref{I_varOmega0}.}
\label{nonfraunh}
\end{figure}

\begin{figure}[thb!]
\includegraphics[width=\columnwidth,clip=true]{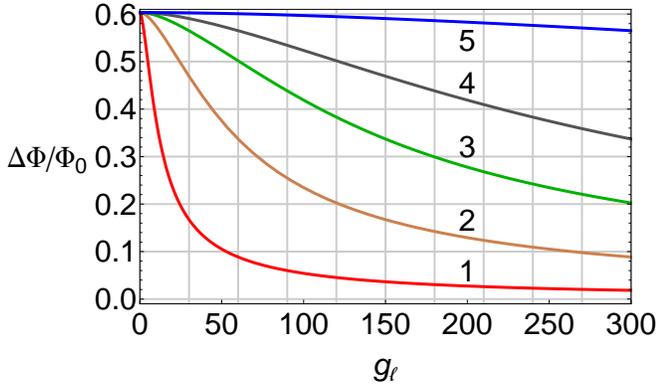}
\caption{The half width at the half of the zero-field peak as a
function of the Josephson coupling constant $g_{\ell}$, according to
equation \eqref{IvarOmega20p}.  The curves are plotted for various
values of the interfacial pair breaking parameter $g_{\delta}$:\,\,
(1)\, $g_{\delta}=4$\,\, (2)\, $g_{\delta}=20$\,\, (3)\,
$g_{\delta}=50$\,\, (4)\, $g_{\delta}=100$\,\, (5)\,
$g_{\delta}=500$.}
\label{width}
\end{figure}

Equations \eqref{IvarOmega03} and \eqref{IvarOmega20p}, as well as
the plots shown in Figs.\,\ref{cprs}, \ref{nonfraunh} and \ref{width}
represent the central results of this paper. Fig.\,\ref{cprs}
displays the current-phase relations obtained numerically with the
exact self-consistent calculations within the GL theory for
$g_\delta=4$ and for various values of $g_{\ell}$ and $\Phi$. The
corresponding equations used for this purpose will be outlined in the
end of the paper. The approximate results based either on equations
\eqref{bcss1012p} and \eqref{I_varOmega0}, or on the analytical
formula \eqref{IvarOmega03}, all lead to the curves, which are in
excellent agreement with and practically indistinguishable from the
exact ones shown in Fig.\,\ref{cprs}. Similarly, the analytical
expression \eqref{IvarOmega20p} agrees well with the numerical
results for the magnetic interference patterns shown in
Fig.\,\ref{nonfraunh}.

The modulations obtained substantially differ from the Fraunhofer pattern
in the junctions with the strongly anharmonic current-phase
dependence, which takes place for large Josephson couplings
$g_{\ell}^2\gg g_{\delta}^2\gg1$. Figs.\,\ref{1},\,\ref{2}
demonstrate that in such junctions the magnetic field substantially
modifies the anharmonic structure of the current-phase relations.
This is in contrast to tunnel junctions (see Fig.\,\ref{0}), where
the sinusoidal shape of the phase dependence is not distorted by
the magnetic field.

An important feature of the non-Fraunhofer patterns, obtained under
the conditions $g_{\ell}\gg g_{\delta}$,\, $g_{\delta}^2\gg 1$, is
that the zero-field peak gets narrower with increasing $g_{\ell}$.
As seen in Fig.\,\ref{width} and as follows from equation
\eqref{IvarOmega20p}, the half width at the half of the peak is small
for large values of $g_{\ell}/g_{\delta}$ and can be approximated
as $(\Delta\Phi/\Phi_0)\approx 1.35 g_{\delta}/g_{\ell}\ll 1$. In
the opposite case $g_{\ell}/g_{\delta}\ll 1$ the quantity
$(\Delta\Phi/\Phi_0)$ is close to its Fraunhofer constant value
$\approx 0.6$ and weakly depends on $g_{\ell}$. Also, the hills at
the finite fields shown in Fig.\,\ref{nonfraunh} are
significantly suppressed as compared to the Fraunhofer ones, in
measure of the same small parameter $g_{\delta}/g_{\ell}\ll1$ and
up to a logarithmic factor. The larger the Josephson coupling, the
stronger the suppression of the critical current by the magnetic
field within each interval $n\Phi_0<\Phi<(n+1)\Phi_0$\,($n=0,\pm1,
\pm2,\dots$). The Fraunhofer pattern is represented by the curve 1
in Fig.\,\ref{nonfraunh}. It appears for $g_\ell\ll
g_{\delta}$, when the first harmonic dominates the current. Then one
gets from \eqref{IvarOmega20p} $I_c(\Phi)=I_{c0}\left|\sin\left(\pi
\Phi\big/\Phi_0\right)\right|\Big/\left(\pi|\Phi|\big/\Phi_0\right)$,
where $I_{c0}=j_{dp}3\sqrt{3}g_{\ell}/4g_{\delta}^2$ is the
corresponding critical current in the absence of the magnetic field.

Let us return now to a brief description of the approach used for plotting
Fig.\,\ref{cprs} and based on the results of Ref. \cite{Barash2012}.
For the order parameter $f(x)e^{i\chi(x)}$ normalized to $f=1$ in the
bulk without superflow, one writes the first integral of the GL
equation in the presence of the supercurrent \cite{Langer1967}
\begin{equation}
\left(\frac{df}{d\tilde{x}}\right)^2+f^2\!-
\frac{1}{2}f^4\!+\,\dfrac{4\tilde{j}^2}{27f^2}=
2f_{\infty}^2-\dfrac{3}{2}f_{\infty}^4.
\label{gl1d8}
\end{equation}
Here $\tilde{x}=x/\xi(T)$, $\tilde{j}={j}/{j_{dp}}=-(3\sqrt{3}/{2})
({d\chi}/{d\tilde{x}} )f^2$ and $f_{\infty}$ is the asymptotic value
of $f$ in the depth of the leads.

The boundary conditions for $f$ and the expression for the
supercurrent in the symmetric junctions are
\begin{equation}
\left(\!\frac{df}{d\tilde{x}}\!\right)_{\pm}=\pm g_b(\chi)f_0, \quad
\tilde{j}=\frac{3\sqrt{3}}{2}g_\ell f_{0}^2\sin\chi.
\label{bcss99}
\end{equation}
One puts $x=0$ in \eqref{gl1d8} and, using \eqref{bcss99}, eliminates
the current and the first derivative of the order parameter. This
results in a biquadratic relation between $f^2_{0}$ and
$f_{\infty}^2$. Then, using the asymptotic formulas in the bulk
$\tilde{j}=(3\sqrt{3}/2)\tilde{v}_s(1-\tilde{v}_s^2)$,\,\,$
f_{\infty}^2=1-\tilde{v}_s^2$ \cite{Tinkham1996} and equating the
current to that in \eqref{bcss99} with
$f_{0}^2=(1-\tilde{v}_s^2)\alpha$, one obtains $\tilde{v}_s=\alpha
g_\ell\sin\chi$. As $\tilde{v}_s$, $f_0$, $f_{\infty}$ and
$\tilde{j}$ are now expressed via the only variable $\alpha$, one
gets from the biquadratic relation the fourth-order polynomial
equation for $\alpha$  \cite{Barash2012}
\begin{equation}
2g_b^{2}(\chi)\alpha-
(1-\alpha)^2[1-\alpha(\alpha+2)g_\ell^{2}\sin^2\chi]=0.
\label{alphaeq}
\end{equation}
Eq.~\eqref{alphaeq} is exact within the conventional GL approach with
the boundary conditions \eqref{bcss99}.  Calculating
$\alpha(g_\ell,g_\delta,\chi)$, one gets $j(\chi)$ and finds
numerically ${\mathit\varOmega}_0(\chi)$.  Then, from equation
\eqref{I_varOmega0} with the calculated ${\mathit\varOmega}_0(\chi)$,
one obtains the self-consistent current-phase relations shown in
Fig.\,\ref{cprs}.

In conclusion, it is demonstrated that the anharmonic phase
dependence of the Josephson current, taking place in the junctions
with a strong Josephson coupling and a pronounced interfacial pair
breaking, has a profound influence on the phase sensitive
magnetic interference patterns. The distinctive features of the
non-Fraunhofer patterns uncovered, are the narrow peak of the critical
current, centered at the zero field, and the suppressed hills at
finite field values. The results obtained could be implemented in
the junctions involving unconventional superconductors and/or
magnetic interfaces.

\begin{acknowledgments}
The support of RFBR grant 11-02-00398 is acknowledged.
\end{acknowledgments}

\providecommand{\noopsort}[1]{}\providecommand{\singleletter}[1]{#1}%

\end{document}